\documentclass[sigconf]{acmart}

\setcopyright{none}
\renewcommand\footnotetextcopyrightpermission[1]{}

\usepackage{fancyhdr}
\fancypagestyle{plain}{%
   \fancyhf{} %
   \fancyfoot[L]{ACM SIGCOMM Computer Communication Review}%
   \fancyfoot[R]{Volume 50 Issue 2, April 2020}%
}
\fancypagestyle{firstpagestyle}{%
   \fancyhf{} %
   \fancyfoot[L]{ACM SIGCOMM Computer Communication Review}%
   \fancyfoot[R]{Volume 50 Issue 2, April 2020}%

}

\usepackage{balance}
\usepackage{subfigure}
\usepackage{booktabs}
\usepackage{pifont}
\usepackage{amsmath}
\usepackage{adjustbox}

\newcommand{\sectionref}[1]{$\S$\ref{#1}}

\begin{document}

\title{The Web is Still Small After More Than a Decade}
\subtitle{A Revisit Study of Web Co-location}

\author{Nguyen Phong Hoang,$^{\dagger}$\textsuperscript{$^*$}\hspace{1em} Arian Akhavan Niaki,$^{\mathsection}$\textsuperscript{$^*$}\hspace{1em} Michalis Polychronakis,$^{\dagger}$\hspace{1em} Phillipa Gill$^{\mathsection}$}
\affiliation{%
 \vspace{-1em}\institution{Stony Brook University, New York, USA$^{\dagger}$ \hspace{0.3em} University of Massachusetts, Amherst, USA$^\mathsection$}
}
\affiliation{%
  \institution{\{nghoang,\hspace{0.2em}mikepo\}@cs.stonybrook.edu \hspace{0.3em} \{arian,\hspace{0.2em}phillipa\}@cs.umass.edu}
}





\begin{abstract}
Understanding web co-location is essential for various reasons. For instance,
it can help one to assess the collateral damage that denial-of-service attacks
or IP-based blocking can cause to the availability of co-located web sites.
However, it has been more than a decade since the first study was conducted in
2007. The Internet infrastructure has changed drastically since then,
necessitating a renewed study to comprehend the nature of web co-location.

In this paper, we conduct an empirical study to revisit web co-location using
datasets collected from active DNS measurements. Our results show that the web
is still small and centralized to a handful of hosting providers. More
specifically, we find that more than 60\% of web sites are co-located with at
least ten other web sites---a group comprising less popular web sites. In
contrast, 17.5\% of mostly popular web sites are served from
their own servers.

Although a high degree of web co-location could make co-hosted sites
vulnerable to DoS attacks, our findings show that it is an increasing trend to
co-host many web sites and serve them from well-provisioned content delivery
networks (CDN) of major providers that provide advanced DoS protection
benefits. Regardless of the high degree of web co-location, our analyses of
popular block lists indicate that IP-based blocking does not cause severe
collateral damage as previously thought.
\end{abstract}

\ccsdesc[500]{Networks~Network measurement}

\keywords{Web co-location, DNS measurement, blocking collateral damage}

\maketitle
\renewcommand{\thefootnote}{\fnsymbol{footnote}}
\footnotetext[1]{Co-first authors}

\section{Introduction}
\label{sec:intro}

Over the last three decades, the World Wide Web (the web for brevity) has
grown exponentially, thanks to the rapid expansion of the Internet. A web site
is a fundamental unit that makes up the web, in which related web resources
(e.g., web pages, images, audios, and videos) are gathered and published via a
web server identified by a domain name (e.g., example.com). Prior to 1997,
each web site was typically hosted on its own server with a distinct IP
address. Therefore, the number of unique IP addresses, with the standard web
port (i.e., 80) open, was an accurate proxy to estimate the number of web
sites at the time. However, since the introduction of name-based virtual
hosting technology as a part of HTTP/1.1 in 1997~\cite{rfc2068}, many web
sites can be co-hosted on the same IP address, making it more challenging and
sophisticated to measure the web, especially in terms of web
co-location~\cite{netcraft_activesites}.

Understanding web site co-location is essential for various reasons. For
instance, it can help one to assess the collateral damage that
denial-of-service (DoS) attacks or IP-based blocking can cause to the
availability of co-located web sites. Shue et al.~\cite{Shue:2007} conducted
the first study of web co-location more than a decade ago and found that the
web was smaller than it seemed in terms of the location of servers. The study
quantifies~i) the extent to which the availability of the web can be affected
by targeted DoS attacks, and~ii) the impact of several IP block lists on
co-hosted web sites. Since then, the Internet has grown dramatically. More
than 354 million domain names have been registered across all top-level domains
(TLDs) as of the second quarter of 2019~\cite{verisign.domains}. In addition,
the adoption of IPv6 and CDNs have changed the way web traffic is delivered.
Considering these drastic changes of the Internet infrastructure over the last
decade, it is desirable to investigate whether previous findings by Shue et
al.~\cite{Shue:2007} still hold in today's web ecosystem.

In this paper, we revisit the study of web site co-location by analyzing
datasets collected from active DNS measurements. Comparing our results with
those of Shue et al.~\cite{Shue:2007}, we find that the web is still small and
centralized to a handful of hosting providers. Some IP addresses of major
hosting providers host from hundreds of thousands to millions of web
sites, which is an increasing trend as these providers often provide web sites
hosted on their infrastructure with not only low-cost DoS protection benefits,
but also access to their well-provisioned CDN. Regardless of the high degree
of web co-location, different from previous observations, our analyses of
popular IP block lists show that their collateral damage is relatively small.
Since these block lists are carefully curated by reputable organizations, a
vast majority of blacklisted IP addresses are associated with only one blocked
domain.

\section{Methodology}
\label{sec:method}

In this section, we review existing DNS measurement methods and discuss the objectives of
our experiment. We also describe how our domain dataset was collected.

\subsection{Existing DNS Measurement Techniques}
\label{sec:dns_measurement}

Using \emph{passive measurement}, DNS data is obtained by an entity who is in a
position to capture DNS traffic from the network infrastructure under its
control (e.g., networks of academic institutes or small
organizations)~\cite{weimer2005passive}. Several previous studies use passive
measurement to observe DNS traffic~\cite{weimer2005passive, Shue:2007,
bilge2011exposure, Kountouras2016, Dell'Amico:ACSAC17}. Passive measurements,
however, may introduce bias in the data collected depending on the time,
location, and demographics of users within the monitored network. Moreover,
another issue with passive data collection is ethics, as data gathered over a
long period of time can reveal online habits of monitored users.

In contrast, \emph{active measurement} involves sending and receiving DNS
queries and responses. Researchers can choose which domains to resolve
depending on the goals of their study, thus having more control over the
collected data. Although this approach can remedy the privacy issue of passive
DNS measurement, it requires an increased amount of resources for running a
dedicated measurement infrastructure if there is a large number of domains
that need to be resolved~\cite{Kountouras2016}. There are prior works that
have been conducting large-scale active DNS measurements for different
purposes and provide their datasets to the community~\cite{Kountouras2016,
rapid7}.

However, these datasets have some specific measurement choices that make them
unsuitable to be used directly for the purpose of our study. First, all DNS
resolutions are issued from a single location (country), while we desire to
observe all potential localized IP addresses due to the deployment of CDNs in
different regions. Moreover, these datasets aim to exhaustively resolve as
many domain names as possible regardless of whether or not they are actively
hosting any web content. We further discuss these differences
in~\sectionref{sub:comparison}.

\subsection{Measurement Objectives}
\label{sec:measurement_objectives}

Although it is desirable for us to resolve all web sites to their IP
address(es), it is incredibly challenging or even unrealistic to resolve all
of them with sufficient regularity (e.g., on a daily basis). As our goal is to
study the nature of web co-location and its impact on web users, it is
reasonable to focus on active sites that are often visited by the majority of
users. To curate such a subset of web sites, we utilized the Alexa and
Majestic lists of site rankings. However, only considering the most popular
web sites would bias our results. Instead, we tried to include as many sites as
possible while keeping our measurements manageable and at the same time,
observing a representative subset of web sites on the Internet.

Due to the increasing adoption of load balancing technologies and CDNs,
exhaustively resolving \emph{all} possible IP addresses for a given domain can
be challenging. To approximate this domain-to-IP mapping, we conducted active
DNS measurements from several vantage points obtained from providers of
Virtual Private Servers (VPS). We tried to select our measurement locations in
a fashion that they are geographically distributed around the globe, thus
allowing us to capture as many localized IP addresses of CDN-hosted domains as
possible. To that end, we choose nine locations for our measurements,
including Brazil, Germany, India, Japan, New Zealand, Singapore, United
Kingdom, United States, and South Africa. Our vantage points span the six most
populous continents.

\subsection{Domain Name Datasets}
\label{sec:domain_name_datasets}

In the original study, Shue et al.~\cite{Shue:2007} conducted analyses on two datasets
of domain names collected from~i) the DMOZ Open Directory Project\footnote{The
authors modify the DMOZ dataset to exclude web sites whose domains are in
the \emph{.com} and
\emph{.net} zone files.} and~ii) the zone files of \emph{.com} and \emph{.net}
TLDs. Although it would be ideal to reproduce the study using similar
datasets, the DMOZ project was closed in 2017. On the other hand, the number
of domains registered under the \emph{.com} and \emph{.net} TLDs has doubled
to 156.1M from 75.7M at the time of the original study in 2007.

Pochat et al.~\cite{LePochat2019} recently propose Tranco, which is a list of
popular domains combined from data of the most recent 30 days of four top
lists that are widely used by the research community: Alexa~\cite{alexa},
Majestic~\cite{majestic}, Umbrella~\cite{cisco_umbrella}, and
Quantcast~\cite{quantcast}. However, each top list has its own pitfalls that
may negatively impact analysis results if used without careful
considerations~\cite{Scheitle:2018:TopList, LePochat2019, Rweyemamu2019}.

To that end, we curated our own domain name dataset from the most recent 30
days of the Alexa and Majestic lists for two reasons. First, these two lists
use ranking techniques that are harder and expensive to
manipulate~\cite{LePochat2019}. Second, they also have the highest number of
common domain names among the four. We do not directly use the Tranco list
since it includes domain names from Quantcast and Umbrella. Particularly,
Quantcast mostly contains sites that are popular only in the
US~\cite{LePochat2019}. Umbrella is highly vulnerable to DNS-based
manipulation~\cite{LePochat2019}, while it also contains many domains that do
not serve web content~\cite{Rweyemamu2019}. 

From each of our VPS locations, we sent iterative A and AAAA queries for 8.6M
fully qualified domain names (FQDNs) on a daily basis. By sending iterative
queries, we make sure that local resolvers are not answering these queries
from their cache, but the answers come from the actual authoritative name
servers. We conducted our measurement for two weeks from July 26th to August
8th, 2019. Our dataset is available at
\textit{\href{https://bit.ly/web-colocation-ccr20}{https://bit.ly/web-colocation-ccr20}}.

For comparison, we also repeat our analysis on two public datasets collected
during the same period provided by the Active DNS
Project~\cite{Kountouras2016} and Rapid7~\cite{rapid7}
in~\sectionref{sub:comparison}. The Active DNS project queries about 242M
domains extracted from zone files of approximately 1.3K TLDs on a daily basis.
Rapid7's dataset consists of a much larger number of domains (2B) obtained
from zone files, web crawling, and domains returned from PTR records by
querying reverse DNS lookup of the whole IPv4 space.

Table~\ref{tab:datasets_stat} summarizes our preliminary observation of unique
domains and IP addresses observed in each dataset. Overall, the numbers of
domains are much larger than the numbers of IP addresses, indicating that
numerous domains are co-hosted under the same IP address(es). We further
analyze this co-location degree in~\sectionref{sec:co-location}
and~\sectionref{sub:comparison}.

\begin{table}[t]
\centering
\caption{Daily breakdown of domains and IP addresses observed from each dataset.}
\begin{tabular}{lrrr}
\toprule
                             & \textbf{VPS Data}  & \textbf{ActiveDNS} &        \textbf{Rapid7} \\ [0.5ex]
\midrule
\small{Unique domains}       &           8.6M &         242M &                        2B \\
\midrule
\small{IPv4-hosted FQDNs}    &           8.2M &         117M &                        1.2B \\
\small{Unique IPv4s}         &           2.1M &        11.5M &                        710M \\
\midrule
\small{IPv6-hosted FQDNs}    &           1.2M &         230K &                         48M \\
\small{Unique IPv6s}         &           280K &          74K &                        8.8M \\
\bottomrule
\end{tabular}
\label{tab:datasets_stat}
\end{table}

\section{Web Co-location Analysis}
\label{sec:co-location}

In this section, we analyze our dataset collected via active DNS measurement
to investigate the extent to which web sites are co-located in terms of IP
addresses and autonomous systems (AS). We also compare our findings with those
found by Shue et al.~\cite{Shue:2007} to examine if previous observations
still apply in today's web ecosystem.

\subsection{Web Server Co-location}
\label{sec:ip-colocation}

The co-location degree can be defined in two ways depending on whether we
consider an IP address or a domain. When an IP address is considered, the
co-location degree is the number of domains hosted on that IP address.
Computing the co-location degree of a given domain, however, is more complex
as a domain can be hosted on several IP addresses. Therefore, we calculate the
co-location degree of a domain by taking the median of co-location degrees
across all IP addresses that host that domain.

\begin{figure}[t]
	\centering
	\includegraphics[width=0.85\columnwidth]{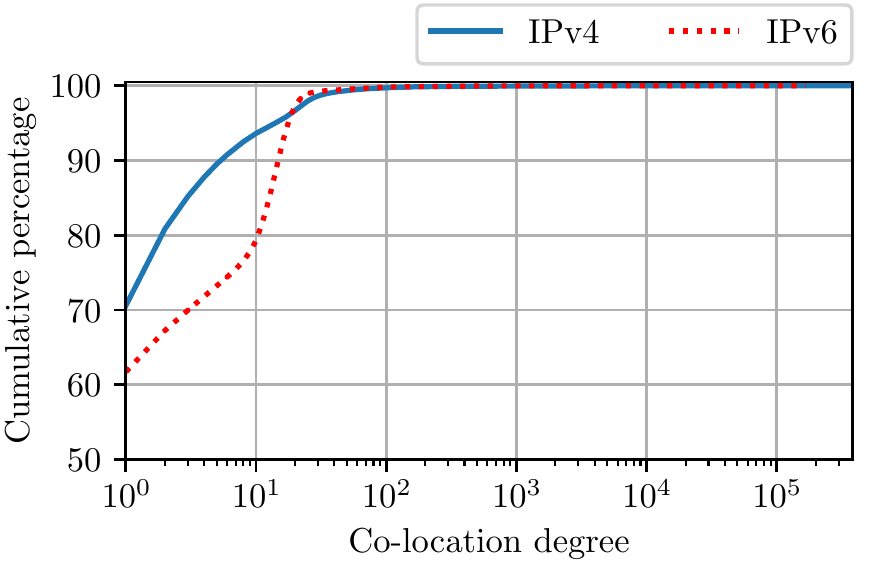}
	\caption{CDF of domains per IP \emph{as a percentage of IPv4/IPv6 addresses} observed in our dataset.}
	\label{fig:web-colocation-as-percent-of-IP}
	\end{figure}

Figure~\ref{fig:web-colocation-as-percent-of-IP} shows the cumulative
distribution function (CDF) of the co-location degree per IP as a percentage
of IPv4/IPv6 addresses observed in our dataset. A large portion of both IPv4
(70.5\%) and IPv6 (61.7\%) are associated with only one domain name. Our
findings are similar to those of Shue et al.~\cite{Shue:2007}, in which 71\%
of the IPv4 addresses in their DMOZ dataset host only one domain.

\begin{figure}[t]
	\centering
	\includegraphics[width=0.85\columnwidth]{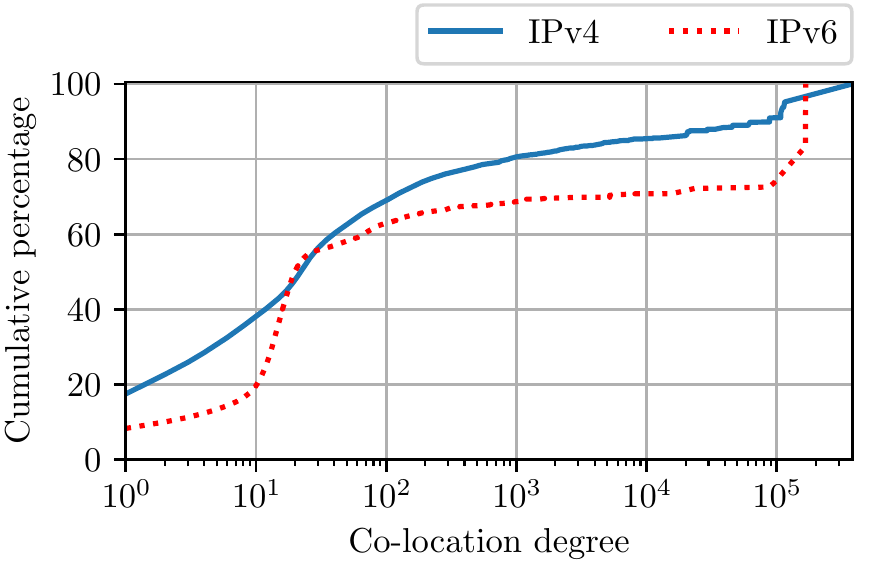}
	\caption{CDF of domains per IP \emph{as a percentage of domains} hosted on IPv4/IPv6 addresses observed in our dataset.}
	\label{fig:web-colocation-as-percent-of-domains}
	\end{figure}

Figure~\ref{fig:web-colocation-as-percent-of-IP} may give an impression that
many domains are hosted on their own IP address since there are many IP
addresses associated with only one domain. However, according to
Figure~\ref{fig:web-colocation-as-percent-of-domains}, there are only 17.5\%
and 8.3\% of domains are hosted on their own IPv4 and IPv6 addresses,
respectively, without sharing the hosting server with any other domain. This
number was higher (24\%) in~\cite{Shue:2007} when considering domains obtained
from the DMOZ dataset.

Figure~\ref{fig:web-colocation-as-percent-of-domains} also indicates that
about 65\% of the web sites in our dataset are co-hosted with 100 or fewer web
sites, decreasing from 84\% in the study of Shue et al.~\cite{Shue:2007}. On
the contrary, we observe that 20\% of domains are co-hosted with more than 1K
other domains on an IPv4 address, increasing from 6\% from the previous
study~\cite{Shue:2007}. Our findings show that more domains are co-hosted
nowadays.

The other end of the CDF in both
Figures~\ref{fig:web-colocation-as-percent-of-IP}
and~\ref{fig:web-colocation-as-percent-of-domains} denote a small number of
IP addresses having an extremely high degree of co-location, hosting a larger
number of domains. The highest co-location degrees are 382K domains for an
IPv4 address, and 167K domains for an IPv6 address. Conducting further
investigation, we find that the IP address with the highest co-location degree
in our dataset belongs to Google, hosting a large number of \emph{blogspot}
sites (i.e., sub-sites of \emph{blogger.com}).

\subsection{Hosting Provider Co-location}
\label{sec:asn-colocation}

Next, we use CAIDA's \emph{pfx2as} dataset~\cite{pfx2as} to map IP addresses
to their organization (i.e., ASN). Similar to the co-location degree of an IP
address (\sectionref{sec:ip-colocation}), the co-location degree in this
section is defined as the number of domain names hosted on the same AS.

\begin{figure}[t]
	\centering
	\includegraphics[width=0.85\columnwidth]{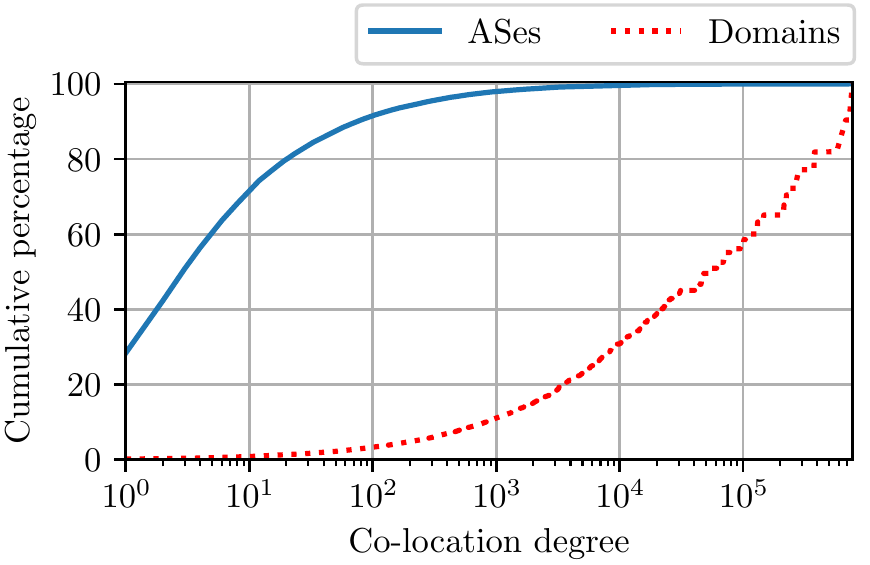}
	\caption{CDF of domains per AS observed in our dataset.}
	\label{fig:web-colocation-as-percent-of-AS}
	\end{figure}

Figure~\ref{fig:web-colocation-as-percent-of-AS} shows the CDF of domains per
AS as a percentage of ASes and domains. 28\% of ASes host only one domain
while only 0.1\% of domains are hosted on an AS themselves. Shue et
al.~\cite{Shue:2007} found that there were 60\% of domains co-hosted with more
than 1K other domains in the same AS. This number has increased to almost 90\%
of domains as indicated in Figure~\ref{fig:web-colocation-as-percent-of-AS}.
These findings again show that an even larger number of domains are co-located
in terms of their hosting provider, indicating that the web is still small
since the first study of Shue et al. conducted more than a decade
ago~\cite{Shue:2007}.

We further analyze our dataset to investigate which organizations dominate
most of the IP addresses and domains. Table~\ref{tab:top_hosting_ASes} shows the
top-ten ASes that~i) occupy the largest portion of IP addresses, and~ii) host
the highest number of web sites. As expected, popular CDN providers (e.g.,
Amazon, Cloudflare, and DigitalOcean) are among the providers from which most
IP addresses were observed. However, in terms of the number of domains,
Cloudflare and Google are the two largest providers, hosting more than 700K
domains each. As Cloudflare provides free web caching services, it is expected
to attract many web owners to proxy their web traffic through Cloudflare's
CDN. While Google is not among the top-ten ASes that dominate the most IP
addresses observed, the company tends to cluster a large number of web sites
under a handful of IP addresses, as shown in~\sectionref{sec:ip-colocation}.
Although a high co-location degree could make co-hosted sites vulnerable to
DoS attacks, our finding shows an increasing trend in which more and more web
sites are co-hosted and served from well-provisioned CDN of major hosting
providers (e.g., Cloudflare, Google). These providers often offer advanced DoS
protection benefits at a relatively low cost~\cite{Cloudflare_dos,
Google_dos}. A higher co-location degree can also potentially improve the
privacy gain of new domain name encryption protocols~\cite{Hoang2020:ASIACCS}.

\begin{table}[t]
\centering
\caption{Top hosting providers that have the highest number of IP addresses/domains.}
\label{tab:top_hosting_ASes}
\setlength{\tabcolsep}{-2pt}
\parbox{.45\linewidth}{
\centering
	\begin{tabular}{lr}
	\toprule
	\textbf{Organization} &\textbf{IPv4s}   \\ [0.5ex]
	\midrule
	AS16509 Amazon        &     130K \\
	AS13335 Cloudflare    &     107K \\
	AS14061 DigitalOcean  &     86K  \\
	AS16276 OVH		      &     76K  \\
	AS46606 Unified Layer &     62K  \\
	AS24940 HETZNER-AS    &     58K  \\
	AS14618 Amazon        &     57K  \\
	AS26496 GoDaddy       &     51K  \\
	AS37963 Alibaba       &     33K  \\
	AS63949 Linode        &     31K  \\
	\bottomrule
	\end{tabular}
}
\phantom{    } 
\parbox{.45\linewidth}{
\centering
	\begin{tabular}{lr}
	\toprule
	\textbf{Organization} &\textbf{Domains}   \\ [0.5ex]
	\midrule
	AS13335 Cloudflare    & 769K \\
	AS15169 Google        & 701K \\
	AS26496 GoDaddy       & 382K \\
	AS46606 Unified Layer & 278K \\
	AS16276 OVH	          & 267K \\
	AS16509 Amazon        & 236K \\
	AS24940 HETZNER-AS    & 229K \\
	AS2635  Automattic    & 145K \\
	AS14061 DigitalOcean  & 130K \\
	AS14618 Amazon        & 129K \\
	\bottomrule
	\end{tabular}
}
\vspace{-4mm}
\end{table}

Many popular web sites are often served from different IP addresses which may
belong to different ASes. We curate our dataset from top lists of popular web
sites, and thus are interested in examining whether these web sites are solely
hosted on one AS or mirrored on several ASes. More specifically, we examine
the top-five populous ASes hosting more than 250K domains, to see if the
domains hosted by them are also hosted on other ASes. If a domain is hosted on
more than one AS, we call it a ``multi-origin'' domain.

\begin{figure}[t]
	\centering
	\includegraphics[width=0.95\columnwidth]{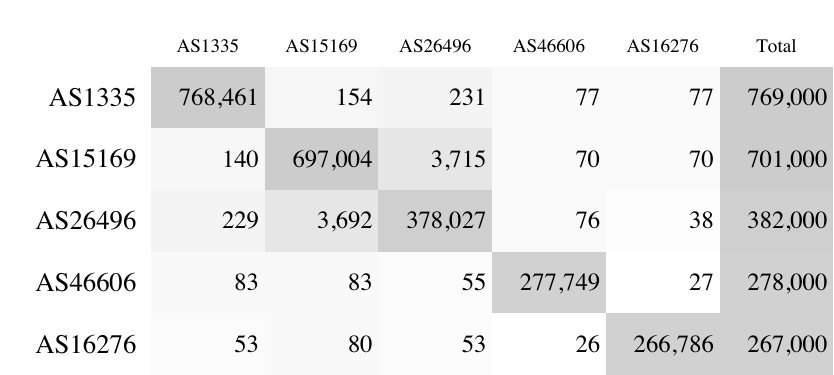}
	\caption{Number of multi-origin domains among top-five autonomous systems. Each cell indicates the number of common domains between two ASes.}
	\label{fig:multi_asn}
	\end{figure}

Although Figure~\ref{fig:multi_asn} shows some multi-origin web sites that are
hosted on more than one AS, the number of such web sites is relatively small.
More than 99.9\% of domains in each AS are only hosted on that AS themselves.
This result again confirms that a vast majority of web sites are centralized
in a handful of hosting providers. Most web sites are hosted solely on one AS
without being mirrored on other ASes. Although most major hosting providers
are equipped with enhanced DoS protections, this single-hosting choice may
have some impact on the availability of web sites hosted on smaller hosting
providers when it comes to targeted DoS attacks.

\vspace{-2mm}
\section{Co-location Degree in Comparison with Larger DNS Datasets}
\label{sub:comparison}

Next, we repeat our analysis conducted in~\sectionref{sec:co-location} using
two larger DNS datasets to examine the extent to which servers are co-located
when considering a much larger number of domain names. More specifically, we
analyze the datasets collected by the Active DNS Project~\cite{Kountouras2016}
and Rapid7~\cite{rapid7} to compare the co-location degree presented in
Figures~\ref{fig:web-colocation-as-percent-of-IP},
\ref{fig:web-colocation-as-percent-of-domains},
and~\ref{fig:web-colocation-as-percent-of-AS} with these two datasets.

\subsection{Dataset Differences}
\label{sec:dataset-differences}

As mentioned in~\sectionref{sec:method}, although the two datasets are similar
to our dataset in terms of measurement methodology (i.e., active measurement),
the location of DNS resolution and the set of resolved domain names are the
two reasons making these datasets unsuitable to be used directly for the
purpose of our study.

With regard to the resolution location, both datasets are collected only from
the US. Particularly, the Active DNS dataset is collected at Georgia Tech
while the Rapid7 dataset is collected from AWS EC2 nodes in the US. This
measurement choice thus could have missed some localized IP addresses of
CDN-hosted domains, which we try to obtain by resolving from multiple
locations in our experiment.

In terms of the number of resolved domain names, both datasets resolve an
order of magnitude larger number of domains than ours as shown in
Table~\ref{tab:datasets_stat}. Most of these domains, however, are not
representative of web sites, while many of them may correspond to spam,
phishing~\cite{Pariwono:ASIACCS18, Tian:imc18, Quinkert:CNS19}, malware
command and control servers~\cite{Alowaisheq:NDSS19}, or parking pages
registered during the domain drop-catching procedure~\cite{Lauinger:usenix17,
Barron:2019:RAID}, which most web users would not typically visit. This is the
primary reason why we opt to curate our own set of domains from the lists of
popular web sites on the Internet.

\subsection{Comparison of Co-location Degree}
\label{sec:colocation-comparison}

\begin{figure}[t]
	\centering
	\includegraphics[width=0.85\columnwidth]{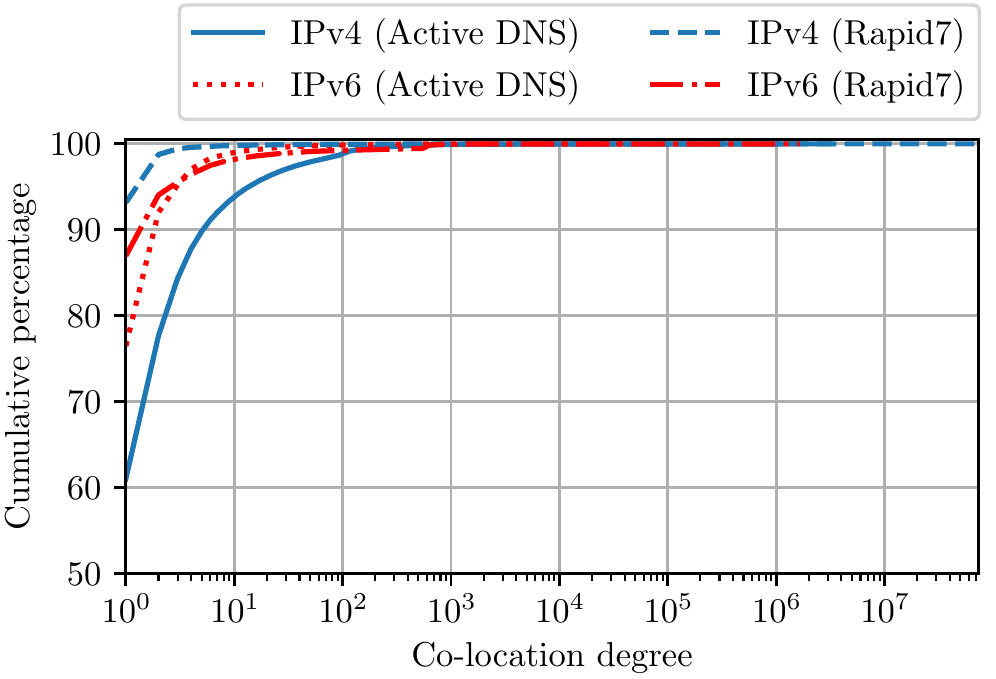}
	\caption{CDF of domains per IP \emph{as a percentage of IPv4/IPv6 addresses} observed in Active DNS and Rapid7 datasets.}
	\label{fig:web-colocation-as-percent-of-IP-comparison}
	\end{figure}

\begin{figure}[t]
	\centering
	\includegraphics[width=0.85\columnwidth]{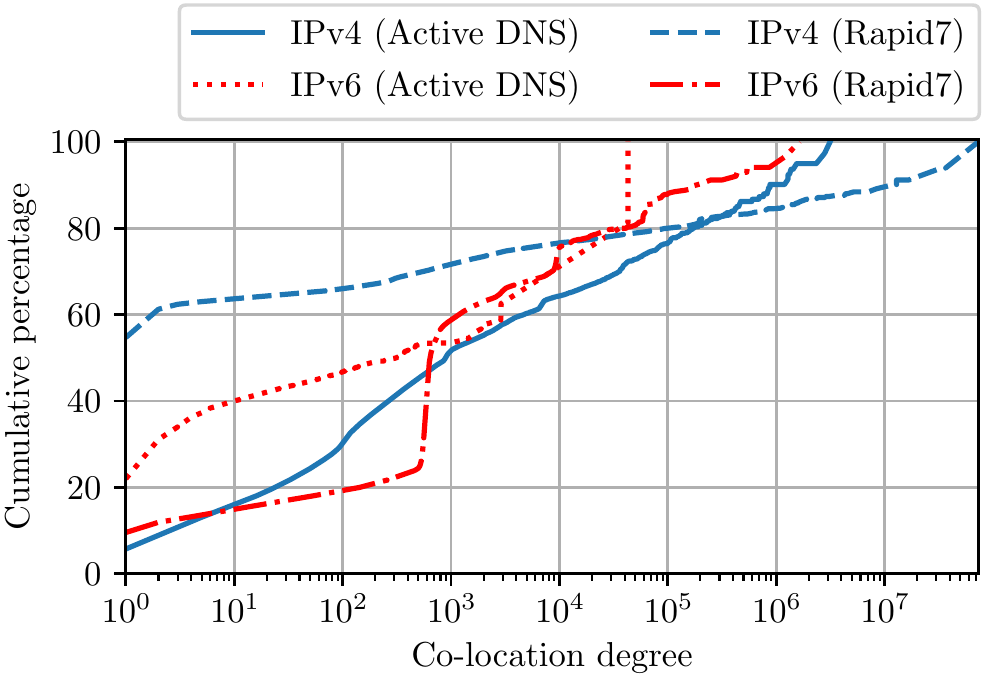}
	\caption{CDF of domains per IP \emph{as a percentage of domains} hosted on IPv4/IPv6 addresses observed in Active DNS and Rapid7 datasets.}
	\label{fig:web-colocation-as-percent-of-domains-comparison}
	\end{figure}

\begin{figure}[t]
	\centering
	\includegraphics[width=0.85\columnwidth]{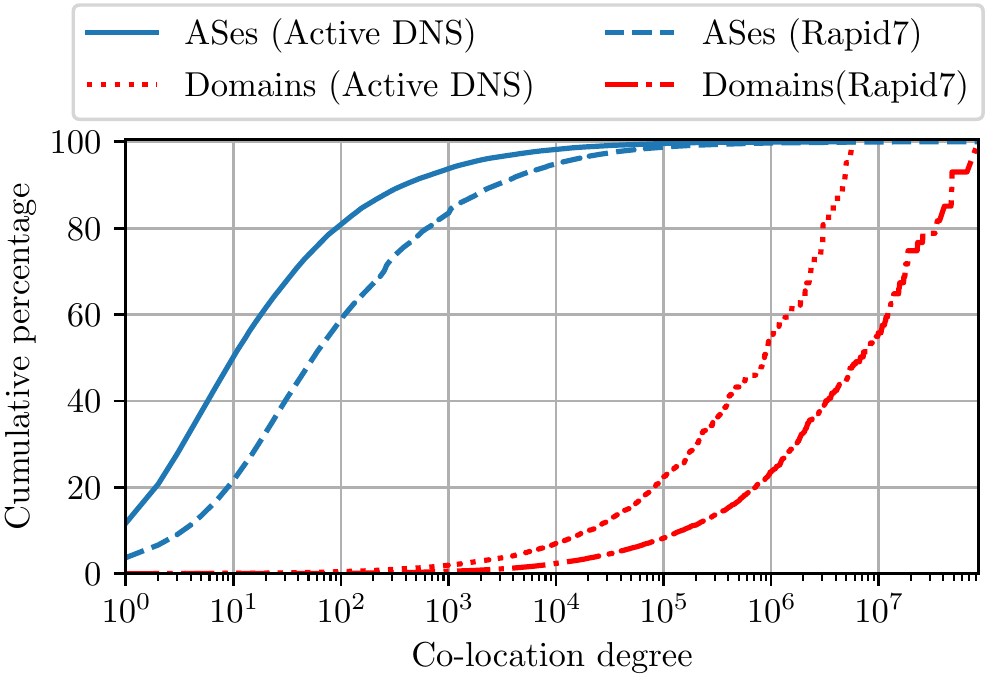}
	\caption{CDF of domains per AS observed in Active DNS and Rapid7 datasets.}
	\label{fig:web-colocation-as-percent-of-AS-comparison}
	\end{figure}

Figure~\ref{fig:web-colocation-as-percent-of-IP-comparison} shows the CDF of
the co-location degree per IP as a percentage of all IPv4/IPv6 addresses
observed in each dataset. Similar to our observation in
Figure~\ref{fig:web-colocation-as-percent-of-IP}, a large number of both IPv4
and IPv6 addresses are associated with only one domain name. More
specifically, 61\% of IPv4 addresses in the Active DNS dataset host only one
domain, while this number is 93\% in the Rapid7 dataset. This result shows a
slight decrease from 69\% of IPv4 addresses that host only one domain observed
in the study of Shue et al.~\cite{Shue:2007} when considering domains obtained
from the \emph{.com} and \emph{.net} zone files.

Similar to our observation in
Figure~\ref{fig:web-colocation-as-percent-of-domains}, the percentage of
domains hosted on an IP themselves is relatively small as shown in
Figure~\ref{fig:web-colocation-as-percent-of-domains-comparison}. The right
end of the CDF denotes domains with an extremely high degree of co-location.
The highest co-location degrees of Active DNS and Rapid7 are 3.1M and 73.3M
per IP address, respectively.

Conducting further investigation, we find that the IP address with the highest
co-location degree in Active DNS dataset belongs to Google Cloud and serves
more than three million personal and small business domains. For Rapid7, the
IP address having the highest co-location degree belongs to AS16276 OVH SAS
and hosts more than 73 million mail servers, instead of web content.
Understandably, a significant portion of domain names used in Rapid7 dataset
consists of PTR records obtained by performing reverse DNS queries over the
whole IPv4 address space. Although most reverse DNS lookups do not return a
meaningful domain name~\cite{Hoang2020:ASIACCS}, an IP address hosting an
email server is required to have a PTR record, storing the domain name of that
email server due to Anti-Spam Recommendations of the Internet Engineering Task
Force~\cite{rfc2505}.

Regardless of resolving a much larger number of domain names,
Figure~\ref{fig:web-colocation-as-percent-of-AS-comparison} shows that only
11.7\% of ASes in the Active DNS dataset and 3.7\% of ASes in the Rapid7 dataset host
one domain, while more than 95\% of domains in both datasets are co-hosted on
the same AS with at least 10K other domains, showing an extremely high level
of AS co-location.

\section{Blocking Collateral Damage}
\label{sub:col-damage}

In this section, we utilize two additional datasets to quantify the collateral
damage on co-located domains of IP-based block lists and censorship-motivated
block lists.

\subsection{IP-based blocking collateral damage}
\label{sec:ip-based-blocking}

\begin{figure*}[t]
	\centering
    \subfigure[VPS Data]{\label{figure:measurement_blocked_cdf}\includegraphics[width=0.33\textwidth,height=0.24\textwidth]{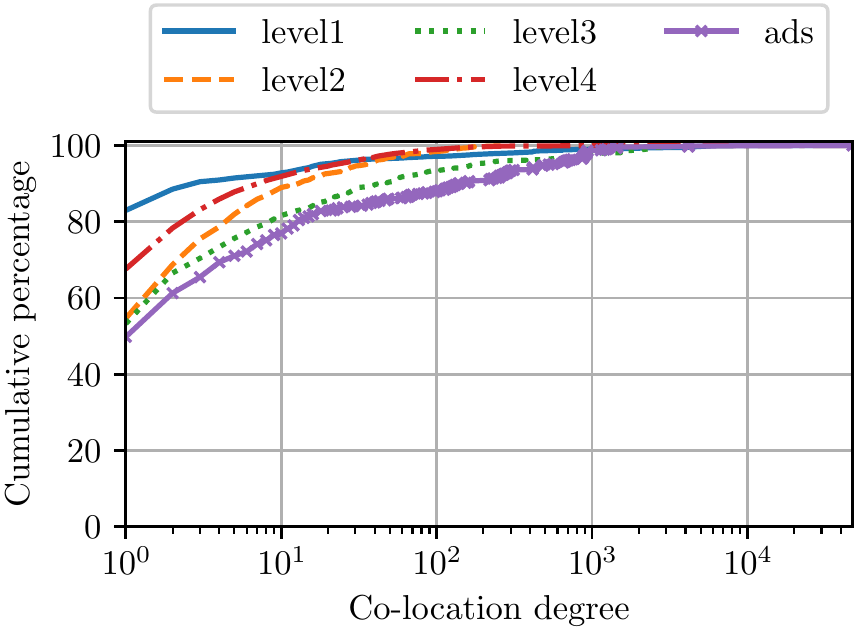}}
    \subfigure[Active DNS]{\label{figure:activedns_blocked_cdf}\includegraphics[width=0.33\textwidth,height=0.24\textwidth]{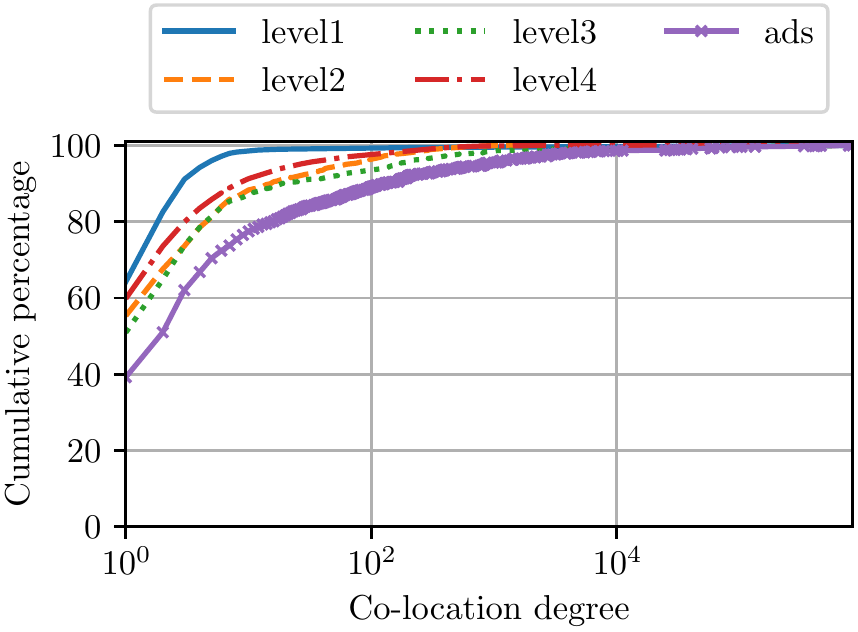}}
    \subfigure[Rapid7]{\label{figure:rapid7_blocked_cdf}\includegraphics[width=0.33\textwidth,height=0.24\textwidth]{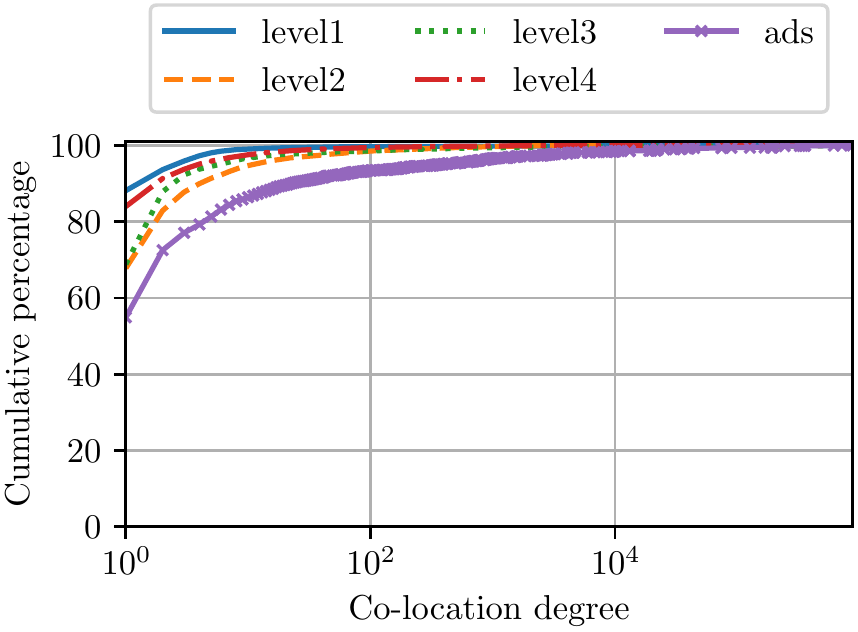}}
	\vspace{-1em}
	\caption{CDF of blocked domains per blacklisted IP address as a percentage
    of common IP addresses between the five block lists and the three DNS
    datasets.}
	\label{figure:blocklists_cdf}
\end{figure*}

In the context of IP-based network filtering, the availability of a web site
can be severely impacted by its co-location degree with other sites. For
instance, if a powerful DoS attack targets a web site, other co-located sites
may also become inaccessible depending on how well-provisioned the hosting
infrastructure is. In the initial study by Shue et al.~\cite{Shue:2007}, to
estimate the collateral damage caused by IP-based blocking, the authors use IP
block lists provided by a security company, which unfortunately no longer
exists.

For our study, we obtained an additional dataset from FireHOL~\cite{firehol},
which is an open-source firewall software that curates its block lists from
several highly reputable sources (e.g., \emph{Abuse.ch}, \emph{DShield.org},
and \emph{Spamhaus.org}). Of its block lists, FireHOL aggregates several
external well-known lists to create four IP block lists, ranked from 1 to 4,
of which \emph{level1} list has minimum false positives, and \emph{level4} may
include a large number of false positives.

More specifically, \emph{level1} is curated to include well-known adversarial
IP addresses monitored by \emph{Spamhaus.org} and \emph{Team-Cymru.org}.
\emph{Level2} includes IP addresses detected to recently conduct brute force
attacks. \emph{Level3} contains malicious IP ranges reported by several
trustworthy sources in the last 30 days. Finally, \emph{Level4} is made from
block lists that track various type of attacks but is susceptible to false
positives. In addition, we also utilize FireHOL's lists that contain IP
addresses of advertising services and trackers.
Table~\ref{tab:blocklist_breakdown} shows the number of IP addresses that each
block list has and the number of common IP addresses found in the three DNS
datasets used in our study. Of these block lists, \emph{Level1} and
\emph{Level4} have the largest numbers of blacklisted IP addresses.

\begin{table}[t]
    \centering
    \small
    \caption{IP addresses in each block list and common IP addresses between
    each list and the three DNS datasets.}
    \begin{tabular}{lrrrr}
    \toprule
    \textbf{Block lists}& \textbf{Unique IPs}  & \textbf{VPS Data}  & \textbf{ActiveDNS} & \textbf{Rapid7} \\ [0.5ex]
    \midrule
    Level1         &   624,564,857   &    3,587       &     34,134         &         145,540 \\
    Level2         &   35,371        &    816         &     1,936          &           8,175 \\
    Level3         &   37,743        &    571         &     1,730          &           9,747 \\
    Level4         &   9,401,369     &    21,224      &     59,948         &         468,026 \\
    Ads            &   13,422        &    595         &     1,895          &           3,594 \\
    \bottomrule
    \end{tabular}
    \label{tab:blocklist_breakdown}
\end{table}

We tested the FireHOL lists, obtained on July 26th, 2019, against the three
DNS datasets to estimate the number of domains affected.
Figure~\ref{figure:blocklists_cdf} depicts the CDF of domains affected per
blacklisted IP address. Across all three DNS datasets, almost 50\% of
blacklisted IP addresses host only one domain. As expected, \emph{Level1} has
the highest percentage of blacklisted IP addresses that host only one domain,
thus causing the least collateral damage. \emph{Level1} is indeed trusted and
widely used by the FireHOL community because the list is carefully compiled
from well-known sources to minimize false positives. Although \emph{level4}
allegedly might include false positives, concerning a high level of collateral
damage, its percentage of blacklisted IP addresses hosting only one domain is
the second-highest (after \emph{Level1}). Overall, less than 10\% of
blacklisted IP addresses of these block lists host more than 100 domains.
Unlike previous observations, this result shows that state-of-the-art IP block
lists are getting better and only cause minimal collateral damage.

\subsection{Censorship collateral damage}
\label{sec:censorship-collateral-damage}

While domain-name-based blocking is one of the dominant techniques that is
often used by censors~\cite{Anonymous:DNS_Damage, china:2014:dns:anonymous,
Wander2014, holdonDNS, farnan2016cn.poisoning, Pearce:2017:Iris,
hoang:2019:measuringI2P}, IP-based blocking can also be very effective for
censorship~\cite{Winter2012a, Arun:foci18, hoang:2018}. Currently, domain name
information is exposed in either DNS queries or the server name indication
(SNI) extension to TLS. This information poses many privacy risks to web users
while making it easier for censors to conduct censorship based on the domain
name. To remedy these problems, new technologies, including DNS over HTTPS/TLS
and ESNI, are introduced to encrypt the domain name information. Under such a
circumstance, censors may shift to IP-based blocking if the domain name
information cannot be obtained.

To quantify the collateral damage of IP-based censorship, we obtained a list
of sensitive sites that are likely to be censored in many countries around the
globe. The list is curated by the Citizen Lab~\cite{citizenlablist} and widely
used in censorship measurement studies~\cite{Filasto2012a, iclab_SP20}. The
list consists of 1,257 web sites, of which we could find 957, 887, and 932
common sites in our dataset, Active DNS, and Rapid7, respectively. We map
these domains to their IP address(es) and investigate how many co-located
domains would be impacted if a censor conducts IP-based filtering to block
these domains.

\begin{figure}[t]
	\centering
	\includegraphics[width=0.85\columnwidth,height=0.52\columnwidth]{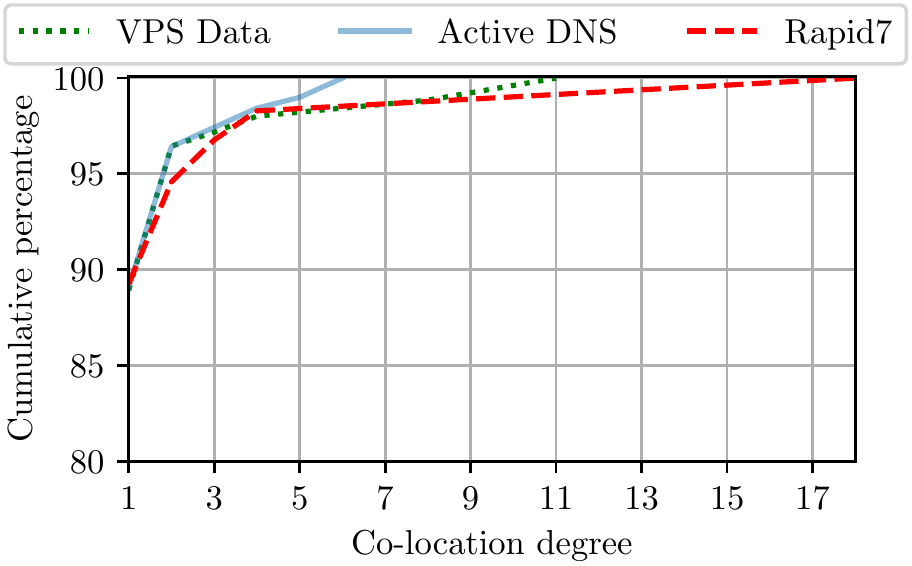}
	\caption{CDF of affected domains per censored IP address as a percentage
	of all observed IP addresses from censored domains in the Citizen Lab
	global sensitive list.}
	\label{fig:citiznlab_us}
	\end{figure}

As indicated in Figure~\ref{fig:citiznlab_us}, nearly 90\% of censored IP
addresses host only one sensitive domain, while the highest numbers of
affected domains are 11, 6, and 18 for our dataset, Active DNS, and Rapid7,
respectively. The result indicates that IP-based censorship will cause little
to no collateral damage. To investigate the reason for this finding, we map
2.8K potentially censored IP addresses to their ASN and find 410 unique
hosting providers. Of these providers, 280 ASes (68\%) host only one domain
while 393 ASes (96\%) host no more than ten sensitive domains from the Citizen
Lab domain list. Therefore, the minimal collateral damage found above is
potentially due to the selection of hosting provider used by censored domains.
On the other hand, previous actions from the side of providers to hinder
domain fronting~\cite{fifield2015blocking, Fahmida2018} have shown that the
collateral damage~\cite{Neil2018} caused to hosting providers may have made
them unwilling to co-host censored domains with other innocuous domains.

\section{Conclusion}
\label{sec:conclusion}

Since its invention, the web has expanded beyond our imagination. More than a
decade ago, Shue et al.~\cite{Shue:2007} conducted the first study of web
co-location and found that the web was smaller than it seemed. In this paper,
we conduct a revisit study of web co-location and could confirm that the web
is indeed still small. More specifically, we find that a large number of web
sites (often less well-known) are co-hosted on a few IP addresses that belong
to major CDN provider. In contrast, a small group of more popular web sites
are served from their own well-provisioned servers, occupying a larger pool of
IP addresses. While this finding of web co-location is similar to its of Shue
et al., our analyses on IP-based blocking show that state-of-the-art IP block
lists are getting better, thus causing a very minimal amount of collateral
damage.
\section*{Acknowledgements}

We are thankful to Manos Antonakakis, Panagiotis Kintis, and Logan O'Hara from
the Active DNS Project for providing us their DNS dataset, and to Rapid7 for
making their datasets available to the research community. We also thank all
the anonymous reviewers for their thorough feedback on earlier drafts of this
paper.

This research was supported in part by the National Science Foundation under
awards CNS-1740895 and CNS-1719386. The opinions in this paper are those of
the authors and do not necessarily reflect the opinions of the sponsors.

{ \balance
{
	\bibliographystyle{ACM-Reference-Format}
	\bibliography{main}
}
}

\end{document}